%% file: Fader.tex
\documentclass{sig-alternate}
\RequirePackage{voila}
\RequirePackage{multirow,tabularx,balance}
\RequirePackage{array}
\begin{document}

\makeatletter
\setlength\textfloatsep{4\p@ \@plus 1\p@ \@minus 1\p@}
\setlength\intextsep   {4\p@ \@plus 1\p@ \@minus 1\p@}
\makeatother

\long\def\symbolfootnote[#1]#2{\begingroup%
\def\thefootnote{\fnsymbol{footnote}}\footnotetext[#1]{#2}\endgroup}

\title{Features and Aggregators for Web-scale Entity Search}
\def\maxauth{2}
\numberofauthors{2}
\author{
Uma Sawant\footnotemark[2]
\and Soumen Chakrabarti\footnotemark[2]
}
\maketitle
\symbolfootnote[2]{IIT~Bombay; contact \url{soumen@cse.iitb.ac.in}}

\input{Abs}

\input{Intro}
\input{Rel}
\input{Feat}
\input{Aggr}
\input{Expt}
\input{End}

\balance
\bibliographystyle{abbrv}
\bibliography{voila}

\end{document}

%% file: Abs.tex
\begin{abstract}
We focus on two research issues in entity search: how to score a document or snippet that potentially supports a candidate entity, and how to aggregate or combine scores from different snippets into an entity score. Proximity scoring has been studied in IR outside the scope of entity search. However, aggregation has been hardwired except in a few cases where probabilistic language models are used. We instead explore simple, robust, discriminative ranking algorithms, with informative snippet features and broad families of aggregation functions. Our first contribution is a study of proximity-cognizant snippet features. In contrast with prior work which uses hardwired ``proximity kernels'' that implement a fixed decay with distance, we present a ``universal'' feature encoding which jointly expresses the perplexity (informativeness) of a query term match and the proximity of the match to the entity mention. Our second contribution is a study of aggregation functions. Rather than train the ranking algorithm on snippets and then aggregate scores, we directly train on entities such that the ranking algorithm takes into account the aggregation function being used. Our third contribution is an extensive Web-scale evaluation of the above algorithms on two data sets having quite different properties and behavior. The first one is the W3C dataset used in TREC-scale enterprise search, with pre-annotated entity mentions. The second is a Web-scale open-domain entity search dataset consisting of 500 million Web pages, which contain about 8 billion token spans annotated automatically with two million entities from 200,000 entity types in Wikipedia. On the TREC dataset, the performance of our system is comparable to the currently prevalent systems by Balog \etal\ (using Boolean associations) and MacDonald \etal. On the much larger and noisier Web dataset, our system delivers significantly better performance than all other systems, with 8\% MAP improvement over the closest competitor.
\end{abstract}

%% file: Intro.tex
\section{Introduction}
\label{sec:Intro}

In its simplest form, entity search queries provide a type (e.g.,
\emph{scientist}) and ask for entities that belong to that type and
satisfy other properties, expressed through keywords (\emph{played
  violin}).  Entity search is a prime example of searching the ``Web
of
Objects''\footnote{\protect\url{http://research.yahoo.com/news/3440}}
or going from ``strings to
things''\footnote{\url{http://googleblog.blogspot.in/2012/05/introducing-knowledge-graph-things-not.html}}
pursued currently by all major search engines.

Machine learning in general, and learning to rank (L2R)
\cite{Liu2009LearningToRank} in particular, can be brought to bear on
entity search in two key interrelated issues:
\begin{itemize}
\item How should a context be scored wrt the query?  Specifically,
  what form of scoring function will take into account the perplexity
  (rarity) of query words and their proximity to (mentions of a)
  candidate entity in a general, trainable fashion?
\item How should evidence from many contexts be aggregated into a
  score or rank for an entity they support?  Can the context scoring
  model be learnt without context labels by directly optimizing for
  entity scores or ranks?
\end{itemize}
Several existing formulations tackle the two issues separately.

\subsection{Proximity scoring}

Scoring documents and passages taking query word match proximity into
account is well established \cite{ButtcherCL2006ProximityBm25,
  TaoZ2007proximity, LvZ2009positional} in IR, but largely outside the
domain of entity search.  Some systems \cite{ChengYC2007EntityRank,
  LiLY2010EntityEngine} use hardwired proximity scoring for entity
search, without using L2R.  Recently, ``proximity kernels'' have been
used \cite{PetkovaBC2007NamedEntityProximity} in entity search based
on generative language models, with tunable width parameters.
However, we know of no end-to-end L2R system where the proximity
scoring function is itself learnt from entity relevance judgments.  As
we shall see here, the issue of robust, trainable proximity scoring is
far from closed.

\subsection{Evidence aggregation}

With very few exceptions~\cite{FangSM2010DiscriminativeExpertSearch,
  MacdonaldO2011RankingAggregates}, entity and expert search
algorithms in the IR community are heavily biased toward generative
language models \cite{CroftL2003lmir, BalogAR2006ExpertSearch,
  FangZ2007ProbabilisticExpertFinding,
  BalogAdR2009LanguageModelExpert}.  In contrast, some of the
best-known L2R algorithms use discriminative max-margin techniques
\cite{HerbrichGO1999ordinal, Joachims2002ranksvm,
  FreundISS2003RankBoost, BurgesSRLDHH2005RankNet,
  BurgesRVL2006LambdaRank, ChapelleLS2007LargeMarginRank,
  YueFRJ2007SVMmap} or conditional probability formulations
\cite{CaoQLTL2007ListNet, VolkovsZ2009BoltzRank,
  ValizadeganJZM2009ndcg}.  In Web search, the best L2R algorithms are
believed to perform considerably better than hardwired scoring
functions from early IR systems.  And yet, entity and expert search
have benefited little from L2R techniques.

A likely reason is the following gap in the respective models.  In
learning to rank (L2R), each item to be ranked is represented by one
feature vector.  In entity search, each item is an entity, potentially
supported by many \emph{contexts}, which may be short token sequences
or entire documents.  Each context, not entity, is associated with a
feature vector.  On the other hand, it is far easier to get entity
relevance judgments than context relevance judgments.

Owing to distributional assumptions, probabilistic retrieval models
\cite{CroftL2003lmir, BalogAR2006ExpertSearch,
  FangZ2007ProbabilisticExpertFinding,
  PetkovaBC2007NamedEntityProximity,
  FangSM2010DiscriminativeExpertSearch} hardwire the manner in which
individual context scores contribute to the score (thereby rank) of an
entity.  As we shall see in Section~\ref{sec:Rel}, these forms of
aggregations have certain limitation.  In later work, Balog
\etal\ \cite{BalogAdR2009LanguageModelExpert} allowed
non-probabilistic aggregations.  Macdonald
\etal\ \cite{MacdonaldO2006VotingExpert,
  MacdonaldO2011RankingAggregates} were the first to systematically
explore a family of aggregation functions and use them as features in
a L2R setting.  They also used hand-crafted rank cutoffs to eliminate
noisy or unreliable support contexts.  Cummins
\etal\ \cite{CumminsLO2010AggregateExpert} used a genetic algorithm to
find a soft rank cutoff.

\subsection{Our contributions}

We started with the goal of unifying hitherto unconnected work on L2R,
proximity scoring and evidence aggregation into a simple and uniform
learning framework.  It turned out that the new framework is also more
robust across diverse data sets, matching or beating all known
systems.

In Section~\ref{sec:Feat} we explore feature design.  In contrast with
earlier proximity kernel \cite{PetkovaBC2007NamedEntityProximity,
  LvZ2009positional} approaches that combine a generative language
model with a decay function having tuned width parameters, we propose
a very general framework for feature design that encodes information
about the rarity (also called ``perplexity'', often measured via
inverse document frequency) of query words matched in a context, as
well as their distance from the candidate entity mention.  In
particular, we do not combine these two signals in a hardwired manner.

In Section~\ref{sec:Aggr} we explore trainable evidence aggregation.
In past work, only Fang
\etal~\cite{FangSM2010DiscriminativeExpertSearch} proposed a document
scoring model that was trained using end-to-end entity relevance
judgment.  We propose a family of pairwise ranking loss
\cite{Liu2009LearningToRank} optimization problems to deal uniformly
with a variety of context score aggregation functions.

In Section~\ref{sec:Expt} we present a detailed experimental study of
the above approaches using two data sets.  The first one is W3C
dataset, from TREC expert search task used in many earlier papers.
This corpus has under 350,000 documents from the W3C Web site with six
different types of web pages (emails, code, wiki, personal homepages,
web and misc).  Since the dataset was used for enterprise search
track, there is only one entity type: \emph{person}.  We performed no
special processing for specific types of pages. The query set for this
dataset contains 50 and 49 ``topics'' from the TREC 2005 and 2006
enterprise tracks.  Relevance judgements were also provided, with
about 4400 relevant candidates for the 99 queries.  To facilitate
standardization, we used the annotated version of W3C dataset
prepared by Jianhan Zhu, available from
\path{https://ir.nist.gov/w3c/contrib/W3Ctagged.html},
containing about 1.6 million annotations.

The second corpus is a representative Web crawl from a commercial
search engine, with 500 million spam-free English documents.  Token
spans that are likely entity mentions are annotated in advance with
IDs from among two million entities belonging to over 200,000 types
from YAGO~\cite{SuchanekKW2007YAGO}.  These annotations (about 8
billion) are then indexed along with text.  We use 845 entity search
queries collected from many years of TREC and INEX competitions,
leading to 93 million contexts supporting candidate entities.  This is
perhaps among the first Web scale entity ranking testbeds where
\emph{all} candidate contexts can be analyzed without depending on a
black-box document-level ranking function with possibly extraneous
scoring considerations like \PageRank\ or click statistics.  We will
place our code and data in the public domain to promote Web-scale
entity ranking research.

\subsection{Results}
\begin{itemize}
\item Purely probabilistic language models that use an
  \emph{expectation} over contexts lose vital signal in $|S_e|$, the
  number of contexts supporting candidate~$e$.

\item However, perplexity+proximity features add further statistically
  significant accuracy to just context count. Very simple features
  that encode perplexity (rarity) of query term matches and their
  proximity from the entity mention are better than fitting proximity
  kernels.

\item On TREC, a simple non-probabilistic sum-of-context-score scheme
  \cite[model~2, Boolean association]{BalogAdR2009LanguageModelExpert}
 and a voting scheme 
  \cite{MacdonaldO2011RankingAggregates,MacdonaldO2006VotingExpert}
are competitive.  However, our system gives comparable performance.

\item On the Web testbed, our system is statistically significantly
  superior to all prior systems.  Thus, the two data sets behave
  differently.  Our system is more robust to the larger corpus with
  noisy entity recognition.
\end{itemize}


%% file: Rel.tex
\section{Related work}
\label{sec:Rel}

We set up some uniform notation.  A query is denoted $q$.  Here we
will model $q$ as a set of words and possibly phrases.  Some of these
may be compulsory for a match, others are optional.  The set of
candidate entities for $q$ is denoted $E_q$, dropping the subscript if
unnecessary.  $e\in E_q$ is a candidate entity (in earlier work
sometimes named $c$ or $\emph{ca}$).

A context supporting a candidate entity may be a whole document or a
short span of tokens (which we call a \emph{snippet}) approximately
centered on a mention of the entity.  An entity may be mentioned in
multiple places in a document.  Likewise, a query term may appear
several times in a document, or even in a snippet.  In this section,
we will use $S_e$ to denote the set of contexts that potentially
support~$e$, without committing on whether $x$ is a document or
snippet.  $x \in S_e$ is one context.

The dominant language modeling approaches 
find the score of context $x$ as $\prod_{t\in q} \Pr(t|x,e)$,
and then aggregate these somehow over $x$ to find a score for~$e$.

\subsection{Scoring one supporting context}
\label{sec:Rel:Prox}

Early expert search systems \cite{BalogAR2006ExpertSearch,
  MacdonaldO2006VotingExpert} did not use proximity signals, and
instead scored the whole supporting document.  Proximity scoring
outside expert search began around the same time
\cite{ButtcherCL2006ProximityBm25, TaoZ2007proximity} or
later~\cite{LvZ2009positional}.  Petkova
\etal\ \cite{PetkovaBC2007NamedEntityProximity} first used proximity
scoring in expert search using ``kernels''.  A proximity kernel
$k(i,o)$ is a non-negative function of a term offset $i$ and an entity
mention offset $o$, that decreases with $|i-o|$.  Instead of using
terms from document $x$ uniformly to construct a language model
$\Pr(t|\theta_x)$, they use $k$ to construct a position-sensitive
language model $\Pr(t|\theta_{x,o})$ where the contribution of the
term $t_i$ at offset $i$ is scaled by $k(i,o)$.  Ranking accuracy is
not very sensitive to the form of $k$; a Gaussian centered at $o$
works well.

\paragraph*{Proximity kernels in generative language models}
Note that, by definition, $\sum_t \Pr(t|\theta_{x,o}) = 1$.  Consider
entities $e_1, e_2$ supported by documents $x_1, x_2$.  In $x_1$,
$e_1$ is mentioned at $o_1=10$, $e_2$ in $x_2$ at $o_2=100$.  Say
there are two query terms, and they occur at positions $8, 13$ in
$x_1$ and $80, 130$ in $x_2$.  Then the models $\Pr(t|\theta_{x,o})$
will be identical for $x=x_1, x_2$, and the absolute proximity
information will be lost.  Based on only $x_1, x_2$, there would be no
reason to prefer $e_1$ over $e_2$.  This is an important limitation of
proximity-based language models that has not been highlighted before,
and that warrants an investigation of purely feature-based approaches,
that we present in Section~\ref{sec:Feat}.

\subsection{Aggregating noisy evidence}

Balog \etal\ \cite{BalogAR2006ExpertSearch,
  BalogAdR2009LanguageModelExpert} were among the first to popularize
generative language models, originally used in traditional IR
\cite{LaffertyZ2001risk}, to expert search.  
Their best model (which we call Balog2)
proceeds as $\Pr(q|e) = \sum_{x\in S_e} \Pr(q|x,e) \Pr(x|e)$.
This leads to a \textbf{sum-product} form:
\begin{align}
\Pr(q|e) = \textstyle
\sum_{x\in S_e} \left(\prod_{t\in q} \Pr(t|x,e) \right) 
\underline{\Pr(x|e)}.
\tag{SumProd}
\label{eq:SumProduct}
\end{align}
The event space associated with $\Pr(x|e)$ has been somewhat
murky; in particular, if an estimate is used such that $\sum_x
\Pr(x|e)=1$, \eqref{eq:SumProduct} effectively becomes a weighted
average or expectation over support documents.

Later, Balog \etal~\cite{BalogAdR2009LanguageModelExpert} proposed a
non-probabilistic scoring scheme by assuming uniform priors over
documents and entities:
\begin{align*}
  \Pr(q|e) &\approx \textstyle \Pr(q|x) \frac{\Pr(e|x)\Pr(x)}{\Pr(e)}\\
&= \frac{1}{|S_e|} \frac{1}{|E_q|} \sum_{x\in S_e} \Pr(q|x) 
\underline{\Pr(e|x)}.
\end{align*}
and then simply omitting the division by $|S_e|$, 
effectively just adding up
context scores, instead of averaging them.  This retains the signal in
the absolute support $|S_e|$, also highlighted as vital by others
\cite{MacdonaldO2006VotingExpert}.  (Note that $|E_q|$, the number of
candidate entities for query $q$, can be ignored even in a truly
probabilistic framework, as it is fixed for the query.)

Macdonald and Ounis \cite{MacdonaldO2006VotingExpert} provided among
the first systematic studies of a space of possible aggregation
functions in collecting evidence from contexts.  However, the paradigm
was restricted to first computing a (fixed, not learnt) score for each
context, lining up a number of aggregates (such as min, max, sum,
average, etc.) and then learning a linear combination among these.
They did not unify voting with feature-based proximity scoring.

Curiously, Macdonald and Ounis \cite{MacdonaldO2006VotingExpert} found
that the ``ExpCombMNZ'' aggregate feature, defined as
\begin{align}
|S_e|\sum_{x\in S_e} \exp\bigl(\text{\em{score}}(x,q)\bigr),
\tag{ExpCombMNZ} \label{eq:ExpCombMNZ}
\end{align}
consistently performed best.  Here {\em{score}} is any function used for calculating match for document $x$ w.r.t query $q$.  Standard examples of such functions include BM25 \cite{SparckJonesWR2000BM25}  and TFIDF cosine.  This is much more extreme than Balog's sum: large scores, exponentiated, will overwhelm smaller scores, and, instead of dividing by $|S_e|$, we \emph{multiply}.  This effect can also be achieved by a rank cutoff \cite{CumminsLO2010AggregateExpert,
  MacdonaldO2011RankingAggregates} or a soft-OR
aggregation~\cite{LiLY2010EntityEngine}.

\subsection{L2R based on entity relevance}

Fang \etal~\cite{FangSM2010DiscriminativeExpertSearch} propose a
noteworthy exception to the above paradigm: write 
\begin{align}
\Pr(e|q) = \sum_{x\in S_e} \underline{\Pr(x)} \Pr(R_{q,x}=1|q,x)
\Pr(R_{x,e}=1|x,e)
\end{align}
with two hidden Boolean random variables $R_{q,x}, R_{x,e}$.  Now
model each component $\Pr(R_{q,x}=1|q,x)$ and $\Pr(R_{x,e}=1|x,e)$ as
a logistic regression.  The formulation is nice in that it permits
training from labeled entities alone; no labeling of contexts is
needed.  However, this flexibility results in a non-convex learning
problem.  Also, thanks to the $\Pr(x)$ term, the signal in $|S_e|$ is
still lost.  Their loss function is itemwise, not pairwise or listwise
\cite{Liu2009LearningToRank}.  (In contrast, ours is pairwise like
\RankSVM~\cite{Joachims2002ranksvm}.)  Furthermore, Fang \etal\ have
no mechanism to capture proximity through features.

\subsection{Some other related systems} 
\label{sec:Rel:Other}

Some systems for large-scale entity search
\cite{ChengYC2007EntityRank, LiLY2010EntityEngine,
  ChengC2010DualInversion} have been reported in the database
community.  Reminiscent of Macdonald, Cummins and coauthors,
\EntityRank~\cite{ChengYC2007EntityRank} assumes additive aggregation
of the form $\sum_x p(x) \text{score} (e,x,q)$ where $x$ is a page and
$p(x)$ its \PageRank.  Proximity scoring was hardwired.  No learning
was involved.  EntityEngine \cite{LiLY2010EntityEngine} is the only
system to have use a soft-or aggregation, but no feature-based
learning was involved.  None of these systems supported open-domain
entities; the largest number of broad entity types supported was
21~\cite{ChengC2010DualInversion}.

\subsection{Overview of our unified framework}
\label{sec:Rel:Framework}

The above picture is somewhat diverse and chaotic, and our main goal
is to unify all the above efforts in a uniform, trainable
feature-based discriminative ranking framework.

\begin{figure}[ht]
\centering\includegraphics{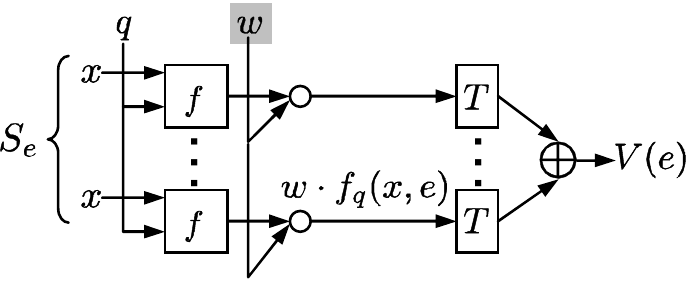}
  \caption{Feature and aggregation formula.}
  \label{fig:Aggr}
\end{figure}

A context $x\in S_e$ has an associated (query dependent) feature
vector $f_q(x,e)\in\R^M$.  
$q$ is dropped if clear from context.  Note that, in general,
$e$ is also an input to $f_q$.  E.g., we may find that features for
\emph{people} should be different from features for \emph{places}.  Or
we may use various collective statistics from $S_e$ inside $f_q$.  
To keep learning simple, we will assume the raw
score of a context is $w \cdot f_q(x,e)$ where $w\in\R^M$ is the
context-level (proximity-cognizant) scoring model to be trained.
Next we must aggregate the raw context scores into a score for $e$:
\begin{align}
V(e)=\bigoplus\{T(w \cdot f_q(x,e)): x \in S_e \}, 
\tag{Aggr} \label{eq:Aggr}
\end{align}
where $\bigoplus$ is a suitable score aggregation operator.  Entities
will be sorted by decreasing $V(e)$ and presented to the user.
\eqref{eq:Aggr} is shown pictorially in Figure~\ref{fig:Aggr}.  Here
$T \in \R\rightarrow\R$ is a (usually monotone) transformation such as
$T(a)=a$, $T(a)=\log(1+a)$, or $T(a)=e^a$.  If $T$ is convex and
fast-growing, we get a soft-max effect, whereas if it is concave
(diminishing returns) we get a soft-count effect.  For some scoring
schemes, $|S_e|$ may be recovered simply by using $T(a) = \llbracket
a>0 \rrbracket$.

Most existing systems can be expressed within the above
paradigm (perhaps with the training part replaced by hand tuning).
E.g., in case of some language models, $w \cdot f_q(x,e)$ can be
interpreted as $w_e \cdot f_q(x)$ where $w_e$ encodes a language model
for $e$ and $f_q(x)$ selects count and position of query words in
context~$x$.


%% file: Feat.tex
\section{Proximity features}
\label{sec:Feat}

In this section we design $f_q(x,e)$ to reward proximity in a
trainable manner.  We will assume the raw score of a context is $w
\cdot f_q(x,e)$ where $w$ is the context-level scoring model to be
trained.  Usually, $f_q(x,e)\ge \vec0$. To the objective functions
that we seek to minimize, we will also add a standard regularization
\cite{Bishop2006PRML} term of the form $\frac{w \cdot w}{2
  \lambda^2}$, where $\lambda$ is a hyperparameter that we fit via
cross validation.

In what follows, the inverse document frequency or $\IDF(t)$ of a term
$t$ is defined as the inverse of the fraction of documents where the
term occurs.  (Instead of the inverse we also tried the negative log
\cite{WittenMB1999giga} but results were not distinguishable.)  This
can also be interpreted as the ``surprise value'' or \emph{perplexity}
of finding $t$ in a context.  Let $\IDF(q) = \sum_{t\in q}
\IDF(t)$.

\subsection{Document vs.\ snippet}

Evidence from different mentions of an entity in a document are
scarcely independent, so it is common \cite{LvZ2009positional} to
choose one mention from each document that is \emph{most favorable} to
the entity, i.e., with the largest score.  Multiple occurrences of
query words in the context offer a similar issue.  When the context is
a short snippet, ignoring all but the match closest to the entity
mention is reported to work well~\cite{ChakrabartiPD2006ir4qa}.

\subsection{Baselines}
\label{sec:Feat:Old}

The \textbf{NoProx} baseline scores the entire document wrt the query
without regard to the position/s of entity mention/s.  TFIDF cosine,
BM25, TFIDF-weighted Jaccard similarity, or probabilistic language
models may be used.  Each such score can be one feature, and $w$ can
combine them suitably.  It is also common to add a constant feature
(value 1, say) which allows $w$ to effectively count the number of
support contexts in $S_e$.

The second baseline, which we expect to be better than the first, is
to use a proximity kernel \cite{PetkovaBC2007NamedEntityProximity}
with a tuned width together with a probabilistic language model.  This
should also approximate well other similar hardwired proximity scoring
schemes \cite{ChengYC2007EntityRank, LiLY2010EntityEngine,
  ChengC2010DualInversion}.  (Note that Lv and Zhai
\cite{LvZ2009positional}, while using a positional language model,
were ranking documents, not entities, so they do not specify any
aggregation logic.)

\subsection{Perplexity-proximity features}
\label{sec:Feat:Grid}

Consider one mention of a candidate entity in a document, together
with just the closest occurrences of each query word that matches in
the document.  Each matched word $t$ is characterized by two
quantities: the perplexity $\IDF(t)$, and the distance\footnote{For
  simplicity we use absolute distance, but left/right can also be
  encoded naturally using signed distance.} $\ell$ (number of tokens)
between the entity mention and $t$.  We now describe three natural
ways to represent the event of $t$ occurring at distance $\ell$ from
the entity mention.

\subsubsection{Cumulative perplexity up to distance $\ell$}

Suppose there are three query terms $q = \{t_1, t_2, t_3\}$, only
$t_2, t_3$ appear in the context, at distances $\ell_2, \ell_3$ from
the mention.  Imagine we are plotting a graph: the x-axis is distance
$\ell$, and the y-axis is the sum of $\IDF(t)/IDF(q)$ for all $t$
matched within distance $\ell$.  The plot starts at $(0,0)$, and jumps
up at any distance where there is a match, in this example,
from $0$ to $\IDF(t_2)/\IDF(q)$ at $\ell_2$, then
to $(\IDF(t_2) + \IDF(t_3))/\IDF(q)$ at $\ell_3$.  The final value is the fraction of query IDF that is matched in the context (1, if all query terms were found).  This forms a normalized feature space for learning $w$.
We call these \textbf{IdfUpto} features.

\subsubsection{Grid features}

\begin{figure}[ht]
\centering\includegraphics{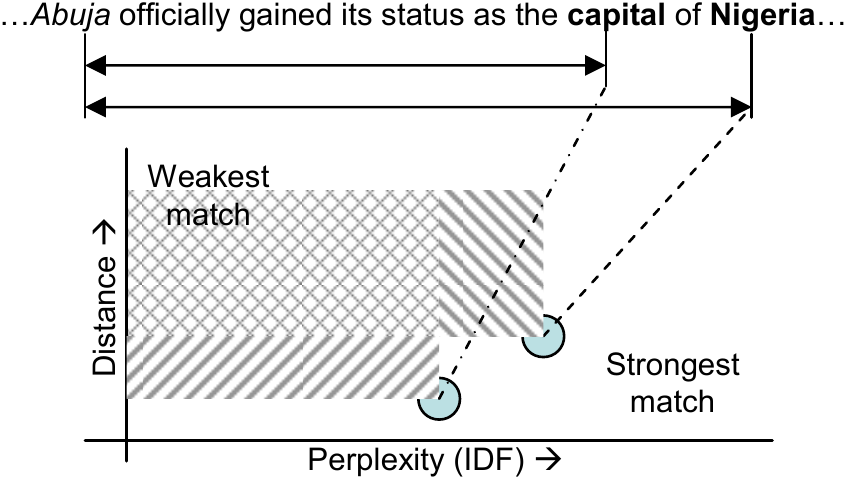}
  \caption{Perplexity-proximity grid features.}
  \label{fig:Grid}
\end{figure}

Now consider a query term $t$ that matches at distance $\ell$ from the
mention of candidate $e$.  Then $\IDF(t)/\IDF(q) \in [0,1]$ decides
the ``perplexity coordinate'' of the match, whereas $\ell$ decides the
``proximity coordinate'' of the match.  In Figure~\ref{fig:Grid},
there are two query terms that match.  \emph{Capital} has lower IDF,
but is closer to the candidate, \emph{Abuja} compared to \emph{Nigeria}, with higher IDF but farther away.

Each axis is suitably bucketed to fire one feature $(i,j)$ in a grid
of features (which is later flattened to a single index in a
1-dimensional vector $f_q(x,e)$).  Every query word match results in
firing one cell in the feature grid (shown as the circles).  This is a
``universal'' encoding, without any commitment on how perplexity and
proximity should be combined; the combination is decided by
learning~$w$.  We call these \textbf{grid features}.

Note that $w$ has an element $w_{i,j} \ge 0$ corresponding to each
grid cell.  Because our discretization is arbitrary, we do not expect
$w_{i,j}$ to differ much from $w_{i\pm1,j\pm1}$.  Therefore, this part
of $w$ should not be regularized (only) as $w_{i,j}^2 / (2\lambda^2)$,
but as
\begin{align*}
 \frac{1}{2\lambda^2} \sum_{i,j} (w_{i,j} - w_{i-1,j})^2
+ (w_{i,j} - w_{i,j-1})^2.
\end{align*}
Assuming row $i+1$ means ``more IDF'' than row $i$ and column $j+1$
means ``more proximity'' than column $j$, we may also want to enforce
monotonicity constraints of the form
\begin{align*}
  w_{i+1,j} \ge w_{i,j} \quad \text{and} \quad w_{i,j+1} \ge w_{i,j}.
\end{align*}

\subsubsection{Rectangle features}

The above forms of constraints over the perplexity-proximity grid
complicate model training, and can be avoided by a
transformation of the grid features to \textbf{rectangle features}.
As Figure~\ref{fig:Grid} shows, each query term matched in the snippet
fires one corresponding grid feature $(i,j)$ (shown by the two circles).  
In the rectangle feature encoding, we also turn on all cells that have
lower IDF or worse proximity.  This ensures that if $(i,j)$ and
$(i',j')$ are close together, the features fired have a large overlap 
(double-hatched area).
Also, the farther to the south-east corner $(i,j)$ is, the more
features are fired.  Note that rectangle features no longer require
the above constraints, just $w_{ij} \ge 0$ is enough.


%% file: Aggr.tex
\section{Evidence aggregation}
\label{sec:Aggr}

As described in Section~\ref{sec:Rel:Framework}, we use a general
expression for entity value \eqref{eq:Aggr} that combines proximity
scoring and evidence aggregation.  The parameters inside
\eqref{eq:Aggr} will be trained using entity-level relevance judgment.

For query $q$ let $G_q, B_q$ be sets of good (relevant) and bad
(irrelevant) entities.  We will use $g, b$ for good and bad entities.
$x_+, x_-$ will denote contexts potentially supporting good and bad
entities.  Before moving on to our suite of aggregation learners, we
note that one may also attempt to directly use L2R techniques at a
context level.  E.g., we can directly use
\RankSVM\ \cite{Joachims2002ranksvm} with hinge loss at the context
level, to minimize wrt $w$ the objective
\begin{align*}
\sum_q  \frac{\displaystyle \sum_{g, b} 
\sum_{\substack{x_+ \in S_g \\ x_- \in S_b}}
\max\{0, 1 + w \cdot \bigl(f_q(x_-,b) - f_q(x_+,b)\bigr)\}}{|G_q||B_q|}
\end{align*}
Note that, while $|G_q||B_q|$ is used to normalize the loss across
queries, the loss is not scaled down by $|S_g|$ or $|S_b|$, which would
average out context support (see Section~\ref{sec:Aggr:SumAverage}).


Context-level formulations are impractical to train.  In our data set,
cases of $10^{11}$ context pairs $(x_+, x_-)$ are not at all rare.
Even an efficient stochastic gradient or sub-gradient descent method
has no hope of dealing with such scale without extreme sampling.  So
some form of direct aggregation to entity score is essential.

Another problem (further motivating Fang \etal's work
\cite{FangSM2010DiscriminativeExpertSearch}) is that a context
supporting a good entity is not necessarily an evidence context as
judged by a human; the entity mention and some query terms may be
juxtaposed coincidentally.  On the other hand, acquiring context-level
supervision is orders of magnitude more expensive than entity-level
relevance judgment.

\subsection{$|S_e|$ baseline}
\label{sec:Aggr:NumSnippet}

A baseline that is known \cite{MacdonaldO2006VotingExpert} to be very
competitive in our testbed is to ignore the quality of the match
between query and context altogether (given at least one term overlap)
and simply use the number of supporting contexts, i.e., $w=(1)$,
$f_q(x,e)=(1)$, $\bigoplus=\sum$, and so $V(e) = |S_e|$.  All other
aggregations must be compared against this trivial baseline.

\subsection{Sum and average}
\label{sec:Aggr:SumAverage}

If we believe $x$ has any signal, and all entity ranking systems
believe so, we should use nontrivial $f_q(x,e)$s.  Two obvious
aggregators that suggest themselves are:
\begin{align}
&& V(e) &= \sum_{x\in S_e} w \cdot f_q(x,e) \tag{Sum} \label{eq:Sum} \\
\text{and} &&
V(e) &= \frac{1}{|S_e|} 
\sum_{x\in S_e} w \cdot f_q(x,e). \tag{Avg} \label{eq:Avg}
\end{align}
Here $T(a) \propto a$.
\eqref{eq:Sum} mimics Balog's non-probabilistic formulation
\cite{BalogAdR2009LanguageModelExpert}.
\eqref{eq:Avg} is our approximation to the evidence aggregation
done by generative language models \eqref{eq:SumProduct}.
Generally, for the sums above to be meaningful, we want no cancellation of
terms, so we will design $f_q(\cdots) \ge \vec0$ and constrain $w \ge
\vec0$.  (For all existing systems this is the case.)

\subsection{SoftMax}
\label{sec:Aggr:SoftMax}

Within the context of TREC entity search, Macdonald
\etal\ \cite{MacdonaldO2006VotingExpert}, Cummins
\etal~\cite{CumminsLO2010AggregateExpert} and others have noted that
not all $x\in S_e$ should contribute to $V(e)$.  Some of these matches
are high-noise and should be tuned down (over and above a hopefully
low context score itself) or eliminated.  They try to achieve this
effect in two different ways.  Cummins \etal\ implement a soft cutoff
as a weighted sum
\begin{align}
  V(e) = \sum_{x \in S_e} D(x; S_e) w \cdot f_q(x,e),
\tag{SoftCutOff}
\end{align}
where $D(x; S_e)$ is a contribution weight that may depend on, e.g.,
the rank of $x$ within $S_e$.  We present a linear program to
learn $D$ in Section~\ref{sec:Aggr:SoftCutoff}.
Macdonald \etal\ instead favor
high-scoring contexts by formulating
\begin{align}
  V(e) = \sum_{x \in S_e} T\bigl( w \cdot f_q(x,e) \bigr),
\tag{SoftMax} \label{eqn:SoftMax}
\end{align}
where $T(\cdot)$ is a fast-growing function, such as $T(a) = e^a$.  A
few high scoring contexts will tend to dominate $V(e)$, hence the
name.  \eqref{eq:ExpCombMNZ} is even more extreme, effectively it
replaces all scores in $S_e$ by the maximum and adds them up.  We
limit ourselves to $T(a) = e^a$.

\subsection{SoftOr}
\label{sec:Aggr:SoftOr}

Although experience with TREC expert search is favorable, it is not
clear if/why soft-max is a universally superior choice.  If a few
high-quality evidence contexts should override other
supporting context scores, another natural aggregator readily suggests
itself: the soft-or (used in
EntityEngine~\cite{LiLY2010EntityEngine}).  The premise here is
\begin{align*}
 \Pr(\text{$e$ is good}|S_e) = 1 - \prod_{x\in S_e} \bigl(
1 - \Pr(\text{$x$ is evidence}) \bigr)
\end{align*}
The standard technique \cite{FangSM2010DiscriminativeExpertSearch} to
turn $w\cdot f_q(x,e)$ into a probability is to use the sigmoid
function $T(a) = \sigma(a) = 1/(1+e^{-a})$:
\begin{align*}
  \Pr(\text{$x$ is evidence for $e$}) = \sigma(w \cdot f_q(x,e))
\end{align*}
In soft-or, $\bigoplus$ is no longer $\sum$, and we get
\begin{align}
V(e) = 1 - \prod_{x \in S_e} \Bigl(1 - \sigma(w \cdot f_q(x,e)\bigr)\Bigr).
\tag{SoftOr} \label{eq:SoftOr}
\end{align}

\subsection{SoftCount}
\label{sec:Aggr:SoftCount}

In both SoftMax and SoftOr, a few large context scores can override a
number of smaller context scores.  An opposite policy, consistent with
the observation that $|S_e|$ is a good scoring scheme by itself, is
that all supporting contexts have some merit, but there is variation
in their evidence quality.  This suggests we use a \emph{concave} $T$,
with a diminishing return shape, instead of a convex one like
$\exp(\cdot)$.  This implements a form of soft counting.  We
specifically used $T(a) = \log(1+a)$ but experience with other forms
like $T(a) = a^p$ with $0<p\le 1$ were similar.

\subsection{Training $w$ through aggregation $\bigoplus$}
\label{sec:Aggr:TrainingThrough}

The earliest L2R formulations seek to minimize the number of wrongly
ordered entity pairs (``pair swaps'') $g \in G_q, b\in B_q$ such that
$V(b) > V(g)$.  The number of pair swaps is directly related to the
area under the curve (AUC) measure in machine learning, and is also
related to MAP by one-sided bounds \cite{Joachims2002ranksvm}.
Minimizing pair swaps \cite{HerbrichGO1999ordinal,
  Joachims2002ranksvm} is a simple and robust L2R approach that
remains hard to beat.  \RankSVM~\cite{Joachims2002ranksvm} proposed to
train $w$ by minimizing wrt $w$ the pair swap hinge loss
\begin{align}
\sum_q \frac{1}{|G_q||B_q|} \sum_{g, b} \max\{0, 1 + V(b) - V(g)\}.
\tag{HingeLoss} \label{eq:HingeLoss}
\end{align}
We will avoid dual solutions  and use simple gradient-descent
optimizers by replacing the hinge loss $\max\{0, 1 + V(b) - V(g)\}$
with the continuous and differentiable soft hinge loss
\begin{align*}
\text{SH}(1 + V(b) - V(g)), \;\text{where}\; 
\text{SH}(a) &= \log(1 + e^a).
\end{align*}
Note that $\text{SH}'(a) = \frac{e^a}{1+e^a} = \sigma(a)$, the sigmoid
function.  The generic gradient of the above loss wrt $w$ is
\begin{align}
  \sum_q \frac{1}{|G_q||B_q|} \sum_{g, b} 
\sigma(1 + V(b) - V(g)) \left[
\frac{dV(b)}{dw} - \frac{dV(g)}{dw} \right],
\tag{Gradient}
\label{eq:SHgrad}
\end{align}
so all that remains is to plug in $V(e)$ and $dV(e)/dw$ for all the
cases discussed.  When $V(e) = \sum_x T(w \cdot f_q(x,e))$, this is
simply $dV(e)/dw = \sum_x T'(w\cdot f_q(x,e)) f_q(x,e)$.  The
$\bigoplus$=\eqref{eq:SoftOr} case
is not additive, but with a little work we can derive:
\begin{align*}
  \frac{d V(e)}{dw} =
  \sum_{x \in S_e} \frac{f_q(x,e)}{1 + e^{-w\cdot f_q(x,e)}}
  \prod_{x' \in S_e}\left( 1 - \frac{1}{1+e^{-w\cdot f_q(x,e)}} \right).
\end{align*}
The soft hinge objective is a convex optimization only in the case
$\bigoplus=\sum$ and $T(a) = a$.  However, (quasi) Newton optimizers
like LBFGS \cite{LiuN1989lbfgs} tend to behave acceptably well even
when given non-convex problems from this
domain~\cite{LiF1999BfgsNonConvex}.  Extending our approach to using
listwise ranking losses \cite{Liu2009LearningToRank} is a possible
direction for future work.

\subsection{Training a soft cutoff by rank}
\label{sec:Aggr:SoftCutoff}

If we assume that $w$ has been trained and fixed, we can set up a
simple linear program (LP) for implementing the kind of soft cutoff
sought by Macdonalds \etal~\cite{MacdonaldO2011RankingAggregates} and
Cummins \etal~\cite{CumminsLO2010AggregateExpert}.  To make it
convenient to use an LP, we will revert from soft hinge to
\eqref{eq:HingeLoss}.  For good, bad entity pair $g, b$, as in
\RankSVM, define slack variable $H(g,b) \ge 0$, with constraint
\begin{align*}
  \forall g, b: &&
  H(g, b) & \ge 1 + V(b) - V(g);
\end{align*}
the objective to minimize will be $\displaystyle \sum_q
\frac{1}{|G_q||B_q|} \sum_{g, b} H(g,b)$.  The soft cutoff decay will
be modeled using more variables $D(r)$ where $r$ is a rank.  Variables
$D$ are constrained by
\begin{align*}
\forall r: &&
D(r) \ge D(r+1), \quad \text{and} \quad D(r) \ge 0.
\end{align*}
Now that $w$ is fixed, all context scores are also fixed.  Let the
rank of context $x$ within $S_e$ be $r_x$.  Then we have
\begin{align*}
  V(e) = \sum_{x \in S_e} D(r_x) (w \cdot f_q(x,e)),
\end{align*}
which can be used to express $H(g,b)$ directly in terms of $D(\cdot)$
and known constants.  As in support vector machines, to limit the
overfitting powers of $D(\cdot)$, we tack on to the objective a
regularization term of the form $D(0)/\lambda$ where $\lambda$ is a
tuned width parameter in the same sense as $w \cdot w / (2 \lambda^2)$
is used in SVM regularization.  Summarizing, the objective will be
\begin{align*}
\min_{H\ge\vec0, D\ge\vec0} 
\frac{D(0)}{\lambda} + \sum_q \frac{1}{|G_q||B_q|} \sum_{g, b} H(g,b)
\end{align*}
subject to the above constraints.  Different entities will have
diverse $|S_e|$.  To share a decay profile $D(r)$ across these, we
allocated parameters in $D(\cdot)$ for deciles of ranks.  We tried
many other rank bucketing approaches but they did not affect the
results significantly.


%% file: Expt.tex
\newcolumntype{C}{>{\centering\arraybackslash} m{2em} }

\section{Experiments}
\label{sec:Expt}

\subsection{Data sets and statistics}

We use two data sets and tasks.  The first one is the standard TREC
enterprise track expert search task used in most prior work on
expert/entity search.  The corpus has 331,000 documents from W3C Web
site.  Mentions of persons (experts) have been annotated (presumably
with near-perfect accuracy) throughout the corpus.  The total number
of annotations is 1.6 million.  The only type of entity sought is a
person (expert) on a given topic, so queries are just bags of words.
On an average a query involves 680 candidate experts.  Expert labels
(relevant/irrelevant) were provided with the queries.  We chose this
reference corpus to make sure our implementation of reported earlier
systems is faithful, with ranking accuracy scores closely matching
published numbers.

But our real interest is in open-domain Web-scale entity search, in
which, as we shall see, competing systems behave rather differently
compared to TREC.  Our second testbed uses a 500 million-page Web
corpus from a commercial search engine.  Token spans that are likely
entity mentions are annotated ahead of time with IDs from among two
million entities belonging to over 200,000 types from
YAGO~\cite{SuchanekKW2007YAGO}.  About eight billion resulting 
annotations were then indexed along with text.

The next step was to collect queries with relevance-judged entities.
We used 845 queries from many years of TREC and INEX.  The
queries were expressed as natural language questions.  There are two
steps to answering these: identify the answer type from among our
200,000 types, and use (some of) the other query words to probe the
text index.  To isolate these two steps, and to align the task to the
TREC task, we had five people rewrite the query into the two
constituents: the answer type and words/phrases to be matched
literally.  Some examples follow:
\begin{itemize}
\item The original query \emph{What is the name of the vaccine for
  chicken pox} was labeled as seeking an entity of the type
  \verb|wordnet_vaccine_104517535| with one or more of these words
  matched close by: \verb|drug +"chicken pox"| \verb|+vaccine|.
\item Likewise, for the original query \emph{Rotary engines were
  manufactured by which company}, the type sought is
  \verb|wordnet_manufacturer_108060446| and the keyword literals may
  be \verb|company +rotary +engine|.
\end{itemize}
The translation was done by proficient search engine users who use +
and quotes properly.  This is not unfair, because the same queries and
retrieval algorithms are available to all competing algorithms.  The
queries are available for anonymous viewing at
\url{http://goo.gl/T2Kkp}.  The five volunteers also curated positive
and negative entity instances from TREC, INEX and the Web; that data
will also be made available in the public domain.

\begin{figure}[ht]
\centering\includegraphics[scale=.8]{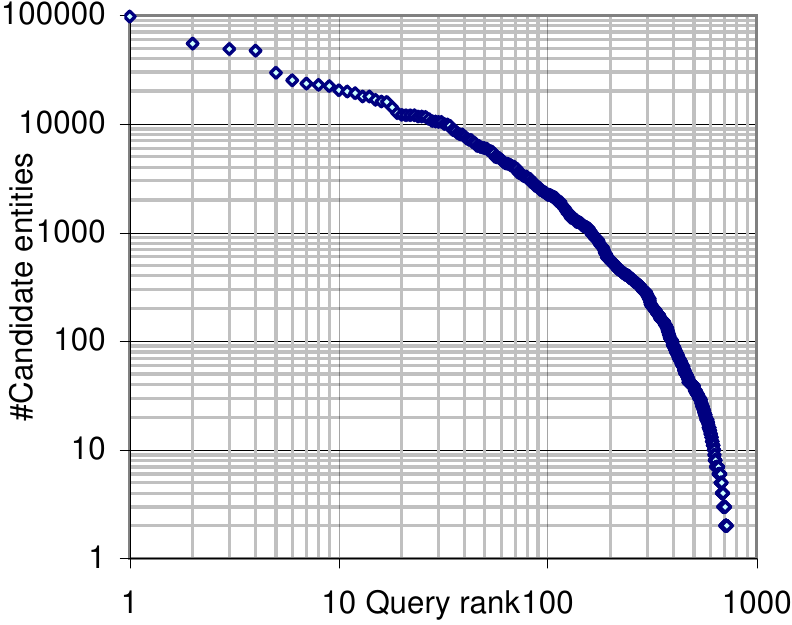}
  \caption{Candidate entities per query.}
  \label{fig:EntsPerQuery}
\end{figure}

\begin{figure}[ht]
\centering\includegraphics[scale=.8]{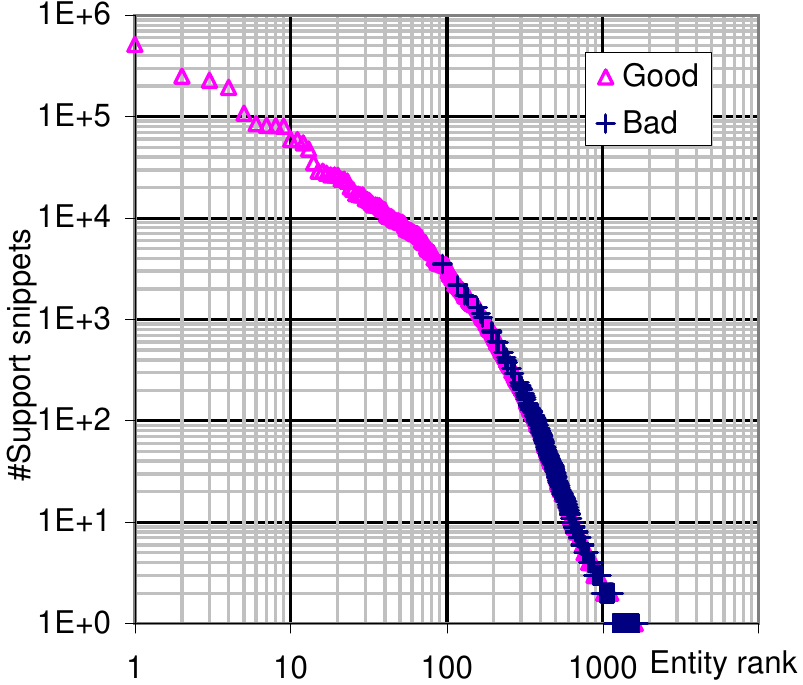}
  \caption{Distribution of supporting contexts ($|S_e|$) per
candidate entity, good and bad.}
  \label{fig:SnipsPerEnt}
\end{figure}

On an average a query leads to evaluating 1884 candidate entities and
110231 contexts, for a total of 93 million contexts over 845 queries.
Figure~\ref{fig:EntsPerQuery} shows the distribution of the number of
candidate entities per query.  Figure~\ref{fig:SnipsPerEnt} shows, for
relevant and irrelevant entities, the number of supporting contexts.
Both plots show heavily skewed behavior.  In particular, from
Figure~\ref{fig:SnipsPerEnt}, we see that some entities are enormously
more popular on the Web compared to others.  Although some good
entities have huge $|S_e|$, we also see that lower down, good and bad
entities are well-mixed in terms of $|S_e|$, and therefore good-bad
separation during ranking remains a challenging problem.

\begin{figure}[ht]
\centering\includegraphics[scale=.8]{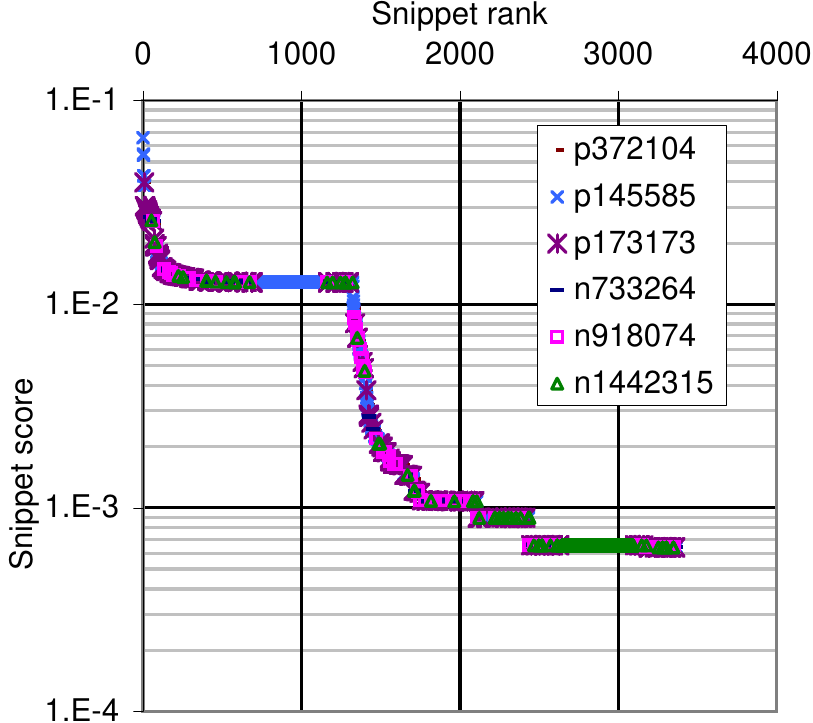}
  \caption{Distribution of context scores in a few $S_e$s
for three good and three bad entities.}
  \label{fig:SnippetScores}
\end{figure}

Figure~\ref{fig:SnippetScores} 
shows context score distributions within some sampled $S_e$s for three
good and three bad entities.  All entities are fairly mixed together
in the chart of context score vs.\ context rank.  Therefore, as with
$|S_e|$ in Figure~\ref{fig:SnipsPerEnt}, entities are not easy to
separate on the basis of score alone.  Three major steps/levels are
seen, corresponding to presence or absence of query keywords.  Within
each broad level, smaller variations near the edges are because of
diverse distances at which query term matches occur.

\subsection{Measurements}
\label{sec:Expt:Measure}

Unless stated, our uniform evaluation policy was \emph{leave one query
  out cross validation}.  We marked each query as the test query, and
trained our parameters on the remaining queries.  Then we evaluated
the trained parameters on the single test query.  Finally we averaged
accuracy measures across all test queries.  This is computationally
intensive, but exploits training data maximally and gives a more
reliable estimate.  In some cases (SoftMax and SoftOr) we reduced the
computational cost of optimization using standard five fold cross
validation across queries.  We report entity level MAP, MRR, NDCG@5,
NDCG@10, and pairs of good/bad entities that are reversed in rank.
The last measure is best if small; others are best if large.

\subsection{Effect of proximity features}
\label{sec:Expt:Prox}

To study the two interacting policies (features and aggregation), here
we will fix the aggregation policy to our overall best (unweighted sum
of context scores, see Section~\ref{sec:Expt:Aggr}), and vary the
design of features.

\begin{figure}[h!]
\begin{center}
\begin{tabular}{|m{5em}|m{2.5em}|m{2.5em}|m{2.5em}|m{2.5em}|m{2.5em}|}
\hline
~& \rotatebox{90}{MAP} & \rotatebox{90}{MRR} & 
\rotatebox{90}{NDCG@5} & \rotatebox{90}{NDCG@10} & \rotatebox{90}{Pairswap}\\
\hline
\raggedright Only $|S_e|$ & $0.542^\downarrow$ & $0.559^\downarrow$ & $0.641^\downarrow$ & $0.660^\downarrow$ & $0.211^\downarrow$ \\
\hline
\raggedright NoProx & 0.559 & $0.578^\downarrow$ & \textbf{0.661} & \textbf{0.675} & 0.203\\
\hline
\raggedright NoProx + IdfUpto & 
0.560 & 0.581 & \textbf{0.661} & 0.672 & $0.221^\downarrow$\\

\hline
\raggedright NoProx + rectangle & \textbf{0.563} & \textbf{0.585} & 0.656 & \textbf{0.675} & \textbf{0.202}\\
\hline
\end{tabular}
\end{center}
\caption{Proximity features compared (Web data).}
\label{fig:Expt:FeatWeb}
\end{figure}

Figure~\ref{fig:Expt:FeatWeb} compares various
proximity features for the Web corpus.  
Rectangle features lead to statistically
significant (paired t-test at $p=0.05$) improvements over not using
proximity signals.  
Here, and in all
tables comparing different settings/systems, in each column, the
largest value is shown in \textbf{boldface}, and other quantities in
the same column that are statistically significantly smaller are
suffixed with a `$^\downarrow$'.  Proximity kernels
\cite{PetkovaBC2007NamedEntityProximity,LvZ2009positional} perform
worse than the numbers in Figure~\ref{fig:Expt:FeatWeb}, but this is
in part due to the way probabilistic language models are built around
the kernels (also see end-to-end comparisons in
Section~\ref{sec:Expt:EndToEnd}).

Figure \ref{fig:Expt:FeatWeb} shows that $|S_e|$ is
already a strong signal for Web data, which concurs with
\cite{MacdonaldO2011RankingAggregates}.  However, the proximity
features bring out significant gains beyond $|S_e|$.

\begin{figure}[h]
\begin{tabular}{|m{5em}|m{2.5em}|m{2.5em}|m{2.5em}|m{2.5em}|m{2.5em}|}
\hline
TREC 05 & \rotatebox{90}{MAP} & \rotatebox{90}{MRR} & 
\rotatebox{90}{NDCG@5} & \rotatebox{90}{NDCG@10} & \rotatebox{90}{Pairswap}\\
\hline
\raggedright Only $|S_e|$ & $0.086^\downarrow$ & $0.271^\downarrow$ & $0.310^\downarrow$ & $0.320^\downarrow$ & $0.459^\downarrow$ \\
\hline
\raggedright NoProx & $0.172^\downarrow$ & $0.511^\downarrow$ & $0.567^\downarrow$ & $0.567^\downarrow$ & 0.371\\
\hline
\raggedright NoProx + IdfUpto & 0.187 & 0.521 & $0.585^\downarrow$ & $0.570^\downarrow$ & 0.373\\
\hline
\raggedright NoProx + rectangle &
\textbf{0.188} & \textbf{0.523} & \textbf{0.606} & \textbf{0.606} &
\textbf{0.370}\\
\hline
\end{tabular}
\begin{tabular}{|m{5em}|m{2.5em}|m{2.5em}|m{2.5em}|m{2.5em}|m{2.5em}|}
\hline
TREC 06 & \rotatebox{90}{MAP} & \rotatebox{90}{MRR} & 
\rotatebox{90}{NDCG@5} & \rotatebox{90}{NDCG@10} & \rotatebox{90}{Pairswap}\\
\hline
\raggedright Only $|S_e|$  & $0.286^\downarrow$ & $0.728^\downarrow$ & $0.718^\downarrow$ & $0.715^\downarrow$ & $0.277^\downarrow$ \\
\hline
\raggedright NoProx & 0.468 & 0.897 & 0.900 & 0.877 & \textbf{0.211}\\
\hline
\raggedright NoProx + IdfUpto & $0.459^\downarrow$ & $0.884^\downarrow$ & 0.892 & $0.865^\downarrow$ & $0.222^\downarrow$\\
\hline
\raggedright NoProx + rectangle &
\textbf{0.477} & \textbf{0.909} & \textbf{0.902} & \textbf{0.879} &
\textbf{0.211}\\
\hline
\end{tabular}
  \caption{Proximity features compared (TREC).}
  \label{fig:Expt:TrecFeat}
\end{figure}

Figure~\ref{fig:Expt:TrecFeat} repeats the feature comparison for
TREC.  A prominent observation is that $|S_e|$ is an excellent single
feature for the Web, but not at all for TREC.  Given that TREC-QA/INEX
queries involve entities well-known to the Web, the ``embarrassment of
riches'' represented in 500 million documents ensures that retrieved
contexts are of generally high quality, so just counting them up is not
too bad.  In contrast, in the much smaller TREC corpus, accidental
similarities between the query and the whole document bring in a large
fraction of poor quality contexts.  This is also confirmed by a later
experiment: while the TREC task benefits from rank-based cutoffs, the
Web task does not. 

\begin{figure}[h!]
\centering\includegraphics[width=25em]{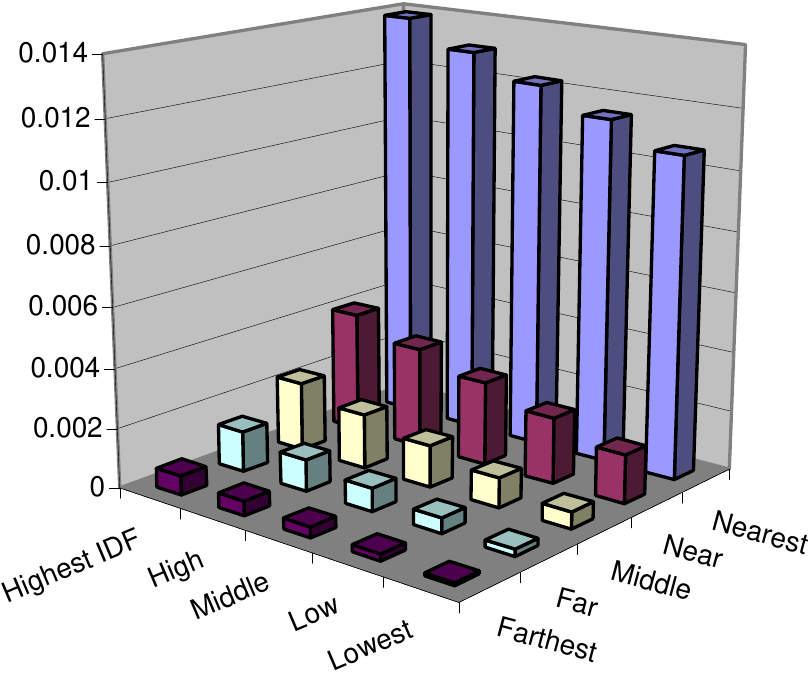}
  \caption{Sample rectangle model weights over the feature grid.}
  \label{fig:Expt:RectangleWeights}
\end{figure}

Figure~\ref{fig:Expt:RectangleWeights} shows the contributions made to
a context score by features firing in each cell shown earlier in
Figure~\ref{fig:Grid}.  If proximity or IDF had no signal, the result
would be a flat-valued weight grid.  Instead we see visible increase
in score contributions as we go from the (low-IDF, large-distance)
corner toward the (high-IDF, small-distance) corner of the grid.

\subsection{Effect of aggregation policies}
\label{sec:Expt:Aggr}

In this subsection we fix the feature representation to the best
reported in the previous subsection, and explore aggregation schemes.
The research questions are:
\begin{itemize}
\item Prior work \cite{CumminsLO2010AggregateExpert,
  MacdonaldO2011RankingAggregates} suggest that contexts in $S_e$
  should not contribute symmetrically to the score of $e$.  Does SoftMax or
  SoftOr perform better than a simple (linear) sum of context scores?
\item Can we get additional mileage beyond linear sum by making
  $T(\cdot)$ sublinear, i.e., using a SoftCount?
\item Can we get improvements by using ranks within $S_e$ to implement
  a soft cutoff (see subsection~\ref{sec:Aggr:SoftCutoff})?
\end{itemize}

\begin{figure}[h!]
\begin{center}
\begin{tabular}{ | m{5em} | m{2.5em} | m{2.5em}| m{2.5em}| m{2.5em}| m{2.5em} | }
\hline
~& \rotatebox{90}{MAP} & \rotatebox{90}{MRR} & 
\rotatebox{90}{NDCG@5} & \rotatebox{90}{NDCG@10} & \rotatebox{90}{Pairswap}\\
\hline
SoftMax & $0.539^\downarrow$ & $0.557^\downarrow$ & $0.639^\downarrow$ & $0.657^\downarrow$ & 0.216\\
SoftOr & $0.357^\downarrow$ & $0.374^\downarrow$ & $0.440^\downarrow$  & $0.460^\downarrow$ & $0.238^\downarrow$\\
SoftCutoff & 0.575 & 0.592 & 0.660 & 0.675 & 0.210\\ 
\hline  \hline
Sum & \textbf{0.576} & \textbf{0.597} & \textbf{0.666} & \textbf{0.681} & \textbf{0.207} \\ \hline
Average & $0.181^\downarrow$ & $0.188^\downarrow$ & $0.227^\downarrow$ & $0.256^\downarrow$ & $0.292^\downarrow$ \\
 \hline \hline
SoftCount & $0.554^\downarrow$ & $0.574^\downarrow$ & 0.661 & 0.675 & 0.207\\
\hline
$|S_e|$  & $0.542^\downarrow$ & $0.559^\downarrow$ & $0.641^\downarrow$ & $0.660^\downarrow$ & 0.211 \\
\hline
\end{tabular}
\end{center}
\caption{Effect of $T, \bigoplus$ and soft cutoffs (Web).}
\label{fig:Expt:TransAggrWeb}
\end{figure}

Figure~\ref{fig:Expt:TransAggrWeb} shows (for Web data) the effect of
choosing score transformer $T$ and aggregator $\bigoplus$ in various
ways for Web data.  Note that $w$ is trained through this choice of
$T, \bigoplus$ as explained in section~\ref{sec:Aggr:TrainingThrough}
and illustrated in Figure~\ref{fig:Aggr}.  The top three rows show the
discriminative aggregation schemes where high-scoring contexts in
$S_e$ get additional preference.  SoftMax uses $\sum_x \exp(w \cdot
f_q(x,e))$, and SoftOr is as described in
subsection~\ref{sec:Aggr:SoftOr}.  SoftCutoff follows
subsection~\ref{sec:Aggr:SoftCutoff}.  The fourth row shows simple sum
$\sum_x w\cdot f_q(x,e)$ with all contexts treated symmetrically.  The
fifth uses \eqref{eq:Avg} instead of sum, and the last two rows show
sublinear aggregation (subsection~\ref{sec:Aggr:SoftCount}) and plain
$|S_e|$ as a trivial baseline.  Figure~\ref{fig:Expt:AggrTrec} shows,
for TREC, the counterpart of Figure~\ref{fig:Expt:TransAggrWeb}.

Linear sum is the clear winner. It is curious that neither superlinear
nor sublinear aggregation beats linear sum.  This could be because
linear sum gives a convex optimization while SoftMax, SoftOr and
SoftCount get trapped in local optima, or because there is something
fundamental about linear sum; this is worthwhile researching further.
Also note that averaging, as against summing, performs poorly,
and $|S_e|$ by itself is not  as good as linear sum.  Even
when scoring was done using linear sum and the SoftCutoff linear
program was used to remove low-scoring contexts' contributions to
entity scores, accuracy dropped.  This lends additional evidence that
symmetric context contribution
to entity score is the best policy.

\begin{figure}
\begin{center}
\begin{tabular}{ | m{5em} | m{2.5em} | m{2.5em}| m{2.5em}| m{2.5em}| m{2.5em} | }
\hline
TREC 05 & \rotatebox{90}{MAP} & \rotatebox{90}{MRR} &
\rotatebox{90}{NDCG@5} & \rotatebox{90}{NDCG@10} & \rotatebox{90}{Pairswap}\\
\hline
SoftMax & $0.072^\downarrow$ & $0.270^\downarrow$ & $0.288^\downarrow$ & $0.294^\downarrow$ & $0.516^\downarrow$\\
SoftOr & $0.075^\downarrow$ & $0.217^\downarrow$ & $0.255^\downarrow$  & $0.299^\downarrow$ & $0.475^\downarrow$\\
SoftCutoff & 0.204 & $0.551^\downarrow$ & $0.400^\downarrow$ & 0.610 & \textbf{0.362}\\               
\hline  \hline
Sum & \textbf{0.207} & \textbf{0.598} & \textbf{0.650} & \textbf{0.614} & 0.369 \\ \hline
Average & $0.111^\downarrow$ & $0.311^\downarrow$ & $0.380^\downarrow$ & $0.372^\downarrow$ & $0.437^\downarrow$ \\
 \hline \hline
SoftCount & 0.195 & $0.524^\downarrow$ & $0.580^\downarrow$ & $0.590^\downarrow$ & 0.372\\
\hline
$|S_e|$  & $0.086^\downarrow$ & $0.271^\downarrow$ & $0.310^\downarrow$ & $0.320^\downarrow$ & $0.459^\downarrow$ \\
\hline
\end{tabular}
\par\smallskip
\begin{tabular}{ | m{5em} | m{2.5em} | m{2.5em}| m{2.5em}| m{2.5em}| m{2.5em} | }
\hline
TREC 06 & \rotatebox{90}{MAP} & \rotatebox{90}{MRR} &
\rotatebox{90}{NDCG@5} & \rotatebox{90}{NDCG@10} & \rotatebox{90}{Pairswap}\\
\hline
SoftMax & $0.257^\downarrow$ & $0.710^\downarrow$ & $0.715^\downarrow$ & $0.684^\downarrow$ & $0.304^\downarrow$\\
SoftOr & $0.214^\downarrow$ & $0.541^\downarrow$ & $0.588^\downarrow$  & $0.610^\downarrow$ & $0.319^\downarrow$\\
SoftCutoff & $0.463^\downarrow$ & $0.891^\downarrow$ & $0.878^\downarrow$ & $0.861^\downarrow$ & 0.210\\
\hline  \hline
Sum & \textbf{0.516} & \textbf{0.933} & \textbf{0.903} & \textbf{0.894} & \textbf{0.203} \\ \hline
Average & $0.148^\downarrow$ & $0.302^\downarrow$ & $0.370^\downarrow$ & $0.379^\downarrow$ & $0.364^\downarrow$ \\
 \hline \hline
SoftCount & $0.456^\downarrow$ & $0.912^\downarrow$ & $0.881^\downarrow$ & $0.855^\downarrow$ & $0.217^\downarrow$\\
\hline
$|S_e|$  & $0.286^\downarrow$ & $0.728^\downarrow$ & $0.718^\downarrow$ & $0.715^\downarrow$ & $0.277^\downarrow$ \\
\hline
\end{tabular}
\end{center}
  \caption{Aggregation choices for TREC.}
  \label{fig:Expt:AggrTrec}
\end{figure}

\subsection{End-to-end comparisons}
\label{sec:Expt:EndToEnd}

Finally, we compare our system's end-to-end accuracy against other
systems, for both data sets.   We compare with these prior systems:
\begin{itemize}
\item Balog2 \cite{BalogAdR2009LanguageModelExpert}, without any
  proximity signal.
\item Macdonald \etal's
  formulation \cite{MacdonaldO2011RankingAggregates} which uses various
  combinations of document scoring models, voting techniques and
  ranking cutoffs.
\item Petkova \etal's formulation using proximity kernel
and generative language model~\cite{PetkovaBC2007NamedEntityProximity}.
\end{itemize}

Although Lv and Zhai \cite{LvZ2009positional} used positional language
models, they did so for document, not entity ranking.  Therefore they
did not specify the all-important aggregation logic needed to turn
their system into an entity search system, and so we cannot directly
compare with them.

Fang \etal's formulation \cite{FangSM2010DiscriminativeExpertSearch}
makes (probabilistic) annotation a query-time activity along with
score aggregation.  While novel, this approach is not practical at Web
scale, where entities may need to be annotated in millions of snippets
at query time.  In both our data sets, annotation is conducted
offline, which effectively turns Fang \etal's system into a single
logistic regression for context scoring, followed by an expectation
over contexts, which we already know as surpassed by $\bigoplus=\sum$.

Petkova \etal\ \cite{PetkovaBC2007NamedEntityProximity} not only
suffer from the same weighted average limitation, but is also
impractical to implement on a Web-scale distributed index.  Instead
of \eqref{eq:SumProduct}, Petkova evaluates the kernel over all
documents for each entity mention, then combines them.  Therefore we
can present numbers for TREC alone.

\begin{figure}[h]
\begin{center}
\begin{tabular}{ | m{5em} | m{2.5em} | m{2.5em}| m{2.5em}| m{2.5em}| m{2.5em} | }
\hline
TREC 05 & \rotatebox{90}{MAP} & \rotatebox{90}{MRR} &
\rotatebox{90}{NDCG@5} & \rotatebox{90}{NDCG@10} & \rotatebox{90}{Pairswap}\\
\hline
Balog2 & 0.214 & 0.604 & $0.623^{\downarrow}$ & 0.617 & $0.347^{\downarrow}$ \\
\hline
Macdonald & \textbf{0.258} & \textbf{0.605} & $0.620^{\downarrow}$ & \textbf{0.623} & \textbf{0.333} \\
\hline
Petkova & $0.101^{\downarrow}$ & $0.370^{\downarrow}$ & $0.371^{\downarrow}$ & $0.373^{\downarrow}$ & $0.549^{\downarrow}$ \\
\hline
Sum & $0.207^{\downarrow}$ & 0.598 & \textbf{0.650} & 0.614 & $0.369^{\downarrow}$ \\  \hline
\end{tabular}
\par\smallskip
\begin{tabular}{ | m{5em} | m{2.5em} | m{2.5em}| m{2.5em}| m{2.5em}| m{2.5em} | }
\hline
TREC 06& \rotatebox{90}{MAP} & \rotatebox{90}{MRR} &
\rotatebox{90}{NDCG@5} & \rotatebox{90}{NDCG@10} & \rotatebox{90}{Pairswap}\\
\hline
Balog2 & \textbf{0.528} & 0.925 & \textbf{0.923} & \textbf{0.921} & \textbf{0.190} \\
\hline
Macdonald & $0.510^{\downarrow}$ & $0.912^{\downarrow}$ & $0.917^{\downarrow}$ & $0.903^{\downarrow}$ & 0.198 \\
\hline
Petkova & $0.318^{\downarrow}$ & $0.731^{\downarrow}$ & $0.747^{\downarrow}$ & $0.751^{\downarrow}$ & $0.332^{\downarrow}$ \\
\hline
Sum & 0.516 & \textbf{0.933} & $0.903^{\downarrow}$ & $0.894^{\downarrow}$ & 0.203 \\ \hline
\end{tabular}
\end{center}
  \caption{End to end comparisons (TREC).}
  \label{fig:Expt:EndToEndTrec}
\end{figure}

Figure~\ref{fig:Expt:EndToEndTrec} shows TREC results.  For TREC 2005,
Macdonald is better than Balog2.  For TREC 2006, Balog2 is better than
Macdonald.  Our system scores slightly less, but is occasionally the
best (NDCG@5 for TREC~05 and MRR for TREC~06).  Petkova implements the
``Balog1'' model \cite[equation~(3)]{BalogAR2006ExpertSearch}, known
for higher recall and lower precision, and falls behind the others.

\begin{figure}[h!]
\begin{center}
\begin{tabular}{ | m{5em} | m{2.5em} | m{2.5em}| m{2.5em}| m{2.5em}| m{2.5em} | }
\hline
~& \rotatebox{90}{MAP} & \rotatebox{90}{MRR} & 
\rotatebox{90}{NDCG@5} & \rotatebox{90}{NDCG@10} & \rotatebox{90}{Pairswap}\\
\hline
Balog2 & $0.452^{\downarrow}$ & $0.468^{\downarrow}$ & $0.543^{\downarrow}$ & $0.565^{\downarrow}$ & $0.173^{\downarrow}$ \\
\hline
Macdonald & $0.535^{\downarrow}$ & $0.553^{\downarrow}$ & $0.616^{\downarrow}$ & $0.634^{\downarrow}$ & $0.222^{\downarrow}$ \\
\hline
Sum & \textbf{0.576} & \textbf{0.597} & \textbf{0.666} & \textbf{0.681} & \textbf{0.207} \\ 
\hline
\end{tabular}
\end{center}
\caption{End to end comparisons (Web).}
\label{fig:Expt:EndToEndWeb}
\end{figure}

Figure~\ref{fig:Expt:EndToEndWeb} is for the Web data set.  Here our
system performs consistently better than all previous approaches
tested, and all differences are significant.  Apart from our parameter
learning, Balog2 does not exploit proximity.  Also, as
Figure~\ref{fig:Expt:TransAggrWeb} (SoftCutoff row) shows, rank-based
cutoffs work worse than symmetric aggregation for the Web, which may
explain why Sum beats Macdonald.


%% file: End.tex
\section{Conclusion}
\label{sec:End}

We presented a system that
unifies diverse, unconnected approaches for context scoring and
entity-level score aggregation into a simple, feature-based,
trainable, discriminative and robust framework for entity ranking.  We
evaluated our system using two data sets.  On the TREC data set, we
are best or close on most evaluation criteria.  On the Web data set,
we are considerably ahead of the competition on all criteria.
The main lessons, some confirming earlier wisdom, were:
\begin{itemize}
\item 
Simple rectangle features, that capture query match perplexity and
lexical proximity, work better than proximity kernels in conjunction
with probabilistic language models.

\item
In case of TREC, $|S_e|$ is a valuable signal; for the Web, it is not.
In all cases, adding more features helped.

\item 
How we aggregate makes or breaks algorithms.  In general, we should
\emph{sum, not average} evidence.  This has serious implications for
probabilistic entity scores that look like $\sum_x (\cdots) \Pr(x|e)$.

\item 
Sublinear (SoftCount) or superlinear (SoftMax) context score
combinations did not yield better ranking than a simple linear
combination; neither did SoftOr.  Rank-based asymmetry in score
aggregation did not help for Web data.
\end{itemize}
Part of our contribution is a fully implemented search system that
answers in a few seconds open-domain entity queries using two million
entities and 200,000 types executed over 500 million Web pages, soon
to be upgraded to two billion pages.  Our code, a demo, and a search
API (as a Web service) will be placed in the public domain.
